\documentclass[twocolumn,preprintnumbers, superscriptaddress, amsmath,amssymb]{revtex4}
\usepackage{epsfig}
\usepackage{subfigure}
\usepackage{amsmath}
\usepackage{amssymb}
\usepackage{graphicx}
\usepackage{dcolumn}
\usepackage{bm}
\input epsf

\begin{document}

\title{Scale dependence of branching in arterial and bronchial trees}

\author{Juan G. Restrepo}
\email{juanga@math.umd.edu}
\affiliation{
Institute for Research in Electronics and Applied Physics,
University of Maryland, College
Park, Maryland 20742, USA
}
\affiliation{
Department of Mathematics,
University of Maryland, College Park, Maryland 20742, USA
}

\author{Edward Ott}
\affiliation{
Institute for Research in Electronics and Applied Physics,
University of Maryland, College
Park, Maryland 20742, USA
}
\affiliation{
Department of Physics and Department of Electrical and Computer Engineering,
University of Maryland, College Park, Maryland 20742, USA
}

\author{Brian R. Hunt}
\affiliation{
Department of Mathematics,
University of Maryland, College Park, Maryland 20742, USA
}
\affiliation{
Institute for Physical Science and Technology,
University of Maryland, College Park, Maryland 20742, USA
}

\date{\today}

\begin{abstract}
Although models of branching in arterial and bronchial trees often
predict a dependence of bifurcation parameters on the scale of the
bifurcating vessels, direct verifications of this dependence by
comparison with data are uncommon. We compare measurements of
bifurcation parameters of airways and arterial trees of different
mammals as a function of scale to general features predicted by
theoretical models based on minimization of pumping power and
network volume. We find that the size dependence is more complex
than existing theories based solely on energy and volume
minimization, and suggest additional factors that may govern the
branching at different scales.
\end{abstract}


\maketitle

The potential factors determining the parameters at bifurcations in
the arterial and bronchial trees of mammals have lately received
attention. Most notably, these local bifurcation characteristics
have been proposed to determine allometric scaling laws for
biologically important variables (see \cite{west,dreyer} and
\cite{rothman}). It has been suggested that arterial and bronchial
trees follow a pattern that minimizes a combination of pumping power
and the total volume of the tree (see for example
\cite{west}-\cite{oka}). If the flow is assumed to be Poiseuille
(resistance proportional to the inverse fourth power of the vessel
radius) on all scales, self-similar branching results, i.e., the
branching parameters do not depend on the scale. However, it has
been argued \cite{west} that, due to the fact that the blood flow is
pulsatile \cite{pulsatile,duan}, arterial trees determined by
optimization have branching parameters that depend on the scale (the
form of this dependence is discussed below). As a direct test of
these theories and their assumptions, we compare in this Letter
experimental measurements of the scale dependence of the branching
parameters with the predictions of theoretical models. Many studies
of bifurcation parameters rely on ordering schemes \cite{horsfield3}
which do not directly consider the dependence of the bifurcation
parameters on the scale, and thus are not appropriate for our
particular purpose. We will explicitly study the dependence of the
radii of the bifurcated vessels on the radius of the parent vessel,
thus allowing a more direct comparison with theoretical models. We
show that important aspects of previous observations
\cite{phillips1} for the human bronchial tree, not predicted by the
theoretical models, are more general than has been previously
noticed. In particular, we find qualitatively similar behavior for
dog arterial and bronchial trees, and lamb fetal and neonatal
arterial trees. Our results suggest that, in addition to minimizing
pumping power and network volume, there might be other important
factors determining the branching in the arterial and bronchial
trees.

Arterial and bronchial trees start with a tree root (a parent
vessel) that divides into two (daughter) vessels, and these in turn
subdivide in a similar fashion, continuing until a certain
approximate value of the radius is attained (in the case of arterial
trees, the capillary radius). At a particular bifurcation, we let
$r_0$ denote the radius of the parent vessel and $r_1$, $r_2$ denote
the radii of the daughter vessels. Variables often used to
characterize the bifurcations are the area ratio and the symmetry
index, defined respectively by $a = (r_{1}^2 + r_2^2)/ r_0^2$ and
$\alpha = r_2/r_1$, where $r_2 \leq r_1$ \cite{zamir2}. These two
variables completely determine $r_1$ and $r_2$ in terms of $r_0$.
The branching exponent $x$ defined for $r_0\geq r_1 \geq r_2$ by
$r_0^x = r_1^x + r_2^x$ is also often used in modeling and
theoretical discussions \cite{woldenberg}-\cite{dawson2}. Most
predictions of theoretical models are stated in terms of this
exponent. However, we find this exponent inconvenient for our
purposes since it diverges as $r_1$ or $r_2$ approach $r_0$, as
occurs frequently with experimental data. We will therefore use the
area ratio in our analysis. The value of the exponent $x$ can be
linked to the change of certain quantities at bifurcations. An
exponent $x = 2$ indicates conservation of total cross sectional
area ($a = 1$) and a value $x = 3$ corresponds to conservation of
laminar shear stress at a bifurcation. The area ratio $a$ is related
to $x$ and the symmetry index $\alpha$ by $a = (1 +
\alpha^2)(1+\alpha^x)^{-\frac{2}{x}}$. This is a monotonic function
of $x$ which does not change much as the value of $\alpha$ is varied
between $0.5$ and $1$. We will therefore regard the general features
of the area ratio as a function of the scale as a counterpart of a
similar qualitative behavior of the exponent $x$.

Murray \cite{murray} proposed that the arterial tree is constructed
in such a way as to minimize a linear combination of the pumping
power and the volume of the network. This leads, if Poiseuille flow
is assumed throughout the tree, to Murray's Law, which states that
$x = 3$. Murray's Law has been useful as a reference point regarding
the small blood vessels and is sometimes mentioned in discussions
regarding the bronchial tree as well \cite{weibel}. However,
deviations have been apparent for some time (see
\cite{kitaoka}-\cite{horsfield} and references therein). The
exponent $x$ has been observed to be generally smaller than the
proposed value of $3$.

Recently, West et al. \cite{west} included the effect of the blood
flow being pulsatile, and found that the minimization scheme carried
on by Murray leads in this case to a transition scenario, with $x$
close to $2$ for the larger vessels and to $3$ for the smaller
vessels (see, however, Ref.~\cite{rothman}). This modification
predicts a smooth transition from the larger to the smaller vessels,
as suggested also by Uylings \cite{uylings} on other grounds. Other
modifications have been proposed predicting an exponent between $2$
and $3$ for the smaller vessels \cite{oka, kassab}. In order to test
these theories we study data for the bifurcation parameter $a$ as a
function of the parent vessel radius. Zamir \cite{zamir2} presents
data for the branching parameters as a function of the radius for
the right coronary artery of a human heart. The large scatter in the
data, however, makes it difficult to observe the dependence of the
mean values on the scale. Phillips and Kaye \cite{phillips1} studied
the dependence of averaged branching parameters on the radius of the
vessels in the bronchial tree of four mammalian species. We show
that some important aspects of their observations for the human
bronchial tree (to be discussed) are more general than has been
previously noticed.

{\bf Data sets.} We study morphological data from the bronchial tree
of a human and 2 dogs obtained by Raabe et al. \cite{raabe}, a human
bronchial tree from Horsfield et al. \cite{horsfield4}, dog lung
arterial trees from Dawson et al. \cite{dawson2} and neonatal and
fetal lamb lung arterial trees from Bennett et al. \cite{bennett}
(see Table~1). For each data set, we compute the average value for
the area ratio $\left<a\right>$ as a function of the parent vessel
radius $r$ as follows. For a given parent vessel radius $r$ we
construct a sample of all bifurcations $(r_0,r_1,r_2)$ which satisfy
$r 10^{-\epsilon} < r_0 < r 10^{\epsilon}$, thus making uniform
intervals in $\log r$, and take $\left<a\right>$ evaluated at $r$ to
be the average of $a$ over this finite sample. We choose the
logarithmic interval width $2\epsilon$ to be $0.05$. This value was
chosen so that the statistical noise is filtered but the large scale
features of $\left<a\right>$ as a function of $r$ are not smoothed
out. For the largest vessels and sometimes for the very small
vessels, the number of bifurcations measured is sparse, especially
when only one lung was measured. We reject the values obtained when
the sample for a given range of radii contains less than $4$
members.
\begin{widetext}
\begin{table}[b]
\centering
\begin{tabular}{|r|r|r|r|r|r|}
\hline
Data set & Mammal & Tree & Body mass (kg) & Radius range (cm) & Bifurcations\\
\hline
Horsfield et al.& Human 1 & Bronchial & - & 0.07 - 1.6 & 3088\\
\hline
Raabe et al.& Human 2 & Bronchial & 81 & 0.04 - 2.01 & 4336\\
\hline
Raabe et al.& Dog 1 & Bronchial & 11.6 & 0.02 - 1.8 & 8032\\
\hline
Raabe et al.& Dog 2 & Bronchial & 10.3 & 0.03 - 1.61 & 2057\\
\hline
Dawson et al.& Dogs (32) & Lung Arterial & 19.7 $\pm$ 2.1 & 0.003 - 0.76 & 919\\
\hline
Bennett et al.& Lamb (fetal)& Lung Arterial & - & 0.001 - 0.70 & 10970\\
\hline
Bennett et al.& Lamb (neonatal)& Lung Arterial & - & 0.011 - 1.23 & 846\\
\hline
\end{tabular}
\caption{Characteristics of the different data sets used. A ``-'' in
the body mass  entry indicates that this information
 was not available}
\label{tab:tabla}
\end{table}
\end{widetext}

\begin{figure}[t]
\begin{center}
\epsfig{file = 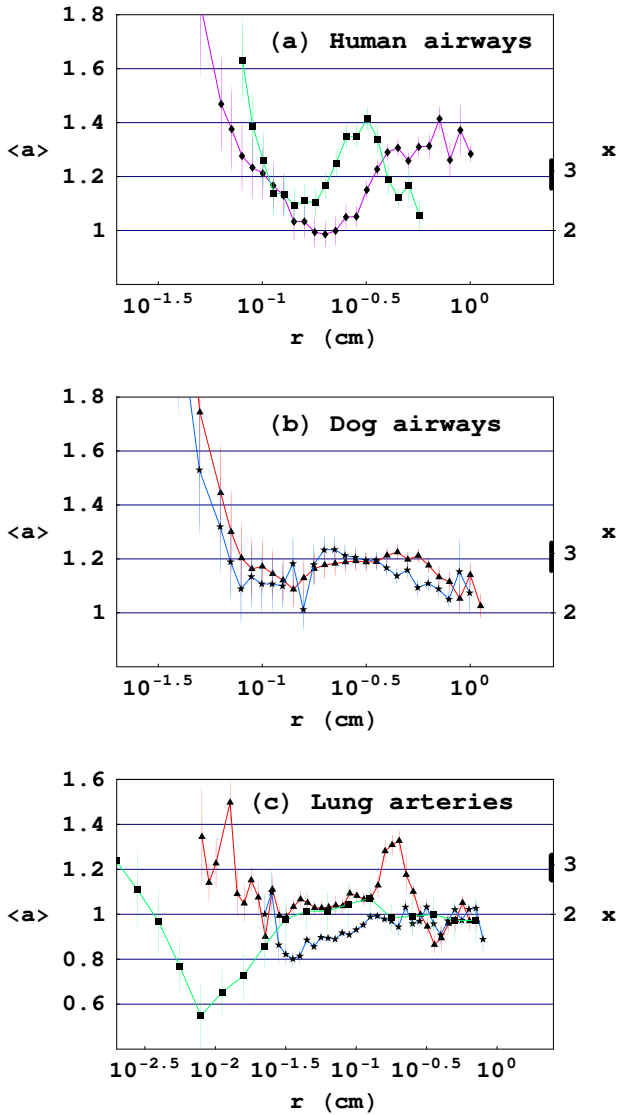, clip =  ,width=1.0\linewidth } \caption{
Average area ratio $\left<a\right>$ as a function of the parent
airway radius $r$ for (a) Horsfield et al. Human 1 (boxes) and Raabe
et al. Human 2 (diamonds), (b) Raabe et al. Dog 1 (triangles) and
Dog 2 (stars) (see Table~1), and (c) Dawson et al. dogs lung
arterial tree (triangles) and Bennett et al. neonatal (stars) and
fetal (diamonds) lambs lung arterial tree.} \label{fig:trifi}
\end{center}
\end{figure}

The results of this procedure are shown in Figure~\ref{fig:trifi}.
In order to allow comparisons with the values of the exponent $x$
predicted by theoretical models, we display the values of $a$
corresponding to $x = 3$ as a thick line on the right axis. The
corresponding value of the area ratio $a$ for $x = 3$ depends on the
symmetry index $\alpha$, and thus we marked the range of values of
$a$ corresponding to $x = 3$ for values of $\alpha$ ranging from
$0.5$ to $1$. The error bars correspond to one standard deviation as
computed from the propagated measurement error taken from the
precision to which the data were reported, and the statistical
uncertainty due to a finite sample.

In Figs.~\ref{fig:trifi}(a) and (b) we show $\left<a\right>$ as a
function of the parent radius $r$ for the human and dog bronchial
trees respectively, obtained  as described above from the Raabe et
al. and Horsfield et al. data sets. In Fig.~\ref{fig:trifi}(c) we
display the area ratio from the dog pulmonary arterial tree, and
fetal and neonatal lamb arterial trees from the data of Dawson et
al. and Bennett et al., respectively. The behavior for the largest
vessels is not shown in the plots because of lack of good
statistics. It has been shown elsewhere \cite{zamir} that the
bifurcations of the largest vessels in the arterial tree generally
have an area ratio close to 1.

We first discuss the results for the dog and human bronchial trees
[Figs.~\ref{fig:trifi}(a) and (b)]. The behavior of $\left<a\right>$
as a function of $r$ is not a simple one: it is not constant, as it
would be if the branching were self-similar, nor does it have a
smooth transition from  a lower value to a higher value as $r$
decreases. Remarkably, the same qualitative behavior is observed in
both graphs. This behavior consists of the following pattern: for
the very small and close to terminal vessels, the area ratio is
high. As $r$ increases, $\left<a\right>$ decreases, reaching a local
minimum at $r \sim 1-2$ mm. Then the area ratio grows, and, for
Human 1 (boxes in [Fig.~\ref{fig:trifi}(a)]) and the dogs
[Fig.~\ref{fig:trifi}(b)], one can observe it finally decreasing
again. In quantitative terms, we make the following observations.
For the humans, among the largest vessels, the value of the area
ratio $\left<a\right>$ peaks higher than that corresponding to a
value of $x = 3$, even when the error bars are taken into account.
For smaller vessels, $\left<a\right>$ decreases as mentioned before,
attaining values below those equivalent to $x = 3$. For the smallest
vessels, the area ratio takes values substantially above those
corresponding to $x = 3$.

We observe a qualitatively similar situation for the lung arteries
[Fig.~\ref{fig:trifi}(c)]. For each data set, there is a local
minimum of the area ratio at moderately small vessel radius, and the
area ratio increases sharply as the radius decreases further. The
fetal and neonatal lamb arterial trees (diamonds and stars,
respectively) have smaller overall values of the area ratio, as
noted before \cite{bennett}. In the dog arterial tree (triangles)
the different regimes are particularly marked: area ratio close to
$1$ for the largest vessels, a peak for the moderately large
vessels, a smaller value of the area ratio  as the radius decreases
further, and larger values of the area ratio near the terminal
vessels. The presence of this peak for moderately large vessels was
noted by Phillips et al. \cite{phillips1} for the bronchial tree of
the human in Raabe et al. data set; we see that this behavior may be
more general.

We also studied data for the airways of two rats and a hamster from
Raabe et al. \cite{raabe}, but the range of vessel radii was smaller
and the results were not conclusive. Data from the heart arteries of
a pig from Kassab et al. \cite{kassab2} also did not show a clear
dependence of area ratio on vessel radius. While not inconsistent
with our observations above, they were not inconsistent with the
hypothesis of a constant area ratio.

The above discussion is qualitative, and it is natural to ask
whether it can be supported by a rigorous statistical test. We
believe, however, that due to the limited amount of samples
available and to the inherent differences between each of the
sampled individuals, we do not have sufficient basis for formulating
an alternative model to those we have discussed. We hope that the
graphical information we have presented will encourage
experimentalists to measure and make available more data.

\begin{figure}[t]
\begin{center}
\epsfig{file = 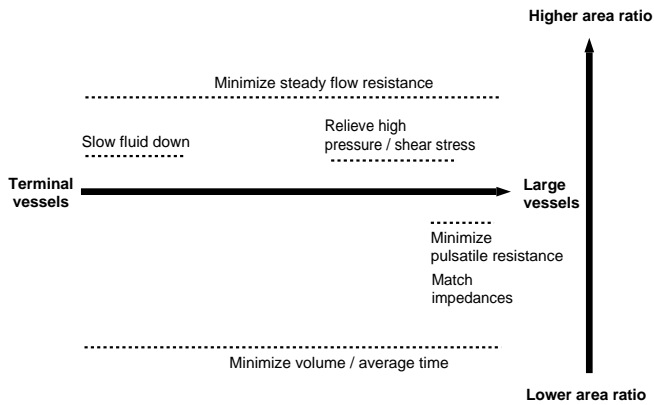, clip =  ,width=1.0\linewidth }
\caption{Schematic representation of possible factors affecting the
area ratio on different scales. The factors above the thick
horizontal line could favor higher area ratios, and those below the
horizontal line favor a lower area ratio (our vertical order is
otherwise arbitrary). A compromise between these competing factors
could produce transitions as seen in Fig.~1.} \label{fig:diagram}
\end{center}
\end{figure}

In summary, a simple transition of the area ratio from values close
to $1$ for the large vessels to higher values for the smaller
vessels, as predicted by some optimization models, is not supported
by the data we studied. The assumption that an optimization
principle is involved in the design of the arterial and bronchial
trees is natural, as evolution would tend to eliminate designs that
are not adapted to meet the demands of the organisms. We believe,
however, that these demands are not restricted to energy efficiency
(as assumed in most theories) and that different factors compete,
resulting in the complex behavior of the area ratio. As has been
remarked recently \cite{mauroy}, a bronchial tree designed solely
with the purpose of minimizing pumping power and network volume
could be dangerously sensitive to deviations from the optimal
bifurcation values, and higher area ratios could be a safety margin
against the effect of these variations. Other factors possibly
contributing to make the area ratio smaller for the largest vessels
are impedance matching and minimization of the resistance for
pulsatile flow \cite{west}. A higher value of the area ratio as the
radius decreases could indicate a tendency to relieve the high
pressure and shear stress present in the large vessels. The large
value of the area ratio near the terminal vessels could allow the
fluid to be slowed down as it reaches the capillaries or alveoli
(e.g., to allow fuller transfer of the transported nutrient to the
body). On the other hand, minimizing the volume or the average
circulation time favors a smaller area ratio on all scales. Also,
the effects of finite vessel length and vessel curvature could
potentially affect the values of the optimal branching parameters.
This discussion is summarized in Fig.~\ref{fig:diagram}. We
emphasize that these are only possible factors, and that this list
is not meant to be exhaustive.
In conclusion, we have presented evidence of a consistent and so far
unexplained behavior in diverse arterial and bronchial trees. To
explain these observations it might be necessary to couple existing
optimality principles to other requirements. The precise formulation
and understanding of such  constraints is a nontrivial and important
problem. The consequences for allometric scaling are also important,
as the branching pattern of the arterial tree has been argued
\cite{west} to be determinant of the scaling of many variables. We
believe it will be fruitful to carry on a similar analysis on other
data.

Acknowledgements: we thank C. A. Dawson and S. H. Bennett for kindly
providing us with data. We also acknowledge the Lovelace Respiratory
Research Institute for allowing us to use their Report LF-53.

\end{document}